\documentclass[%
 reprint,
 amsmath,amssymb,
 aps,
]{revtex4-1}

\usepackage{graphicx}
\usepackage{dcolumn}
\usepackage{bm}
\usepackage{physics,placeins,xcolor}
\usepackage[normalem]{ulem}

\renewcommand{\ol}[1]{\overline{#1}}

\newcommand{\aop}{\hat{a}}
\newcommand{\cop}{{\aop^\dagger}}

\begin{document}

\preprint{APS/123-QED}

\title{Classical critical dynamics in quadratically driven Kerr resonators }

\author{Wouter Verstraelen}
\author{Michiel Wouters}%
\affiliation{%
TQC, University of Antwerp, B-2610 Wilrijk
}%

\date{\today}

\begin{abstract}
Driven-dissipative kerr lattices with two-photon driving are experimentally relevant systems known to exhibit a symmetry-breaking phase transition, which belongs to the universality class of the thermal Ising model for the parameter regime studied here. In this work, we perform finite-size scaling of this system as it is quenched to the transition and the dynamical critical exponent is found to be compatible with $z\approx2.18$ corresponding with metropolis dynamics in classical simulations. Furthermore, we show that the Liouvillian gap scales with the same exponent, similar to scaling of the Hamiltonian gap at quantum phase transitions in closed systems.
\end{abstract}

\maketitle


\section{\label{sec:level1}Introduction}

Thermal phase transitions in classical systems connected to a heath bath,  as well as quantum phase transitions in closed quantum systems, are well-understood \cite{KersonHuang_2014,sachdev}. Very recently, a strong interest has emerged in the critical behavior of open quantum systems \cite{breuer}, for example in a driven-dissipative context, fueled by the realization that such systems may gain importance as a workhorse in quantum technologies \cite{Verstraete2009}.

The central object in such systems is not the free energy as in thermal phase transitions or the Hamiltonian as in quantum phase transitions, but the Liouvillian superoperator \cite{CarmichaelBOOK,liouvillianSpectral}. This object can exhibit a rich phenomenology \cite{Nigro2019}: as both the quantum phase transition and the thermal phase transitions are --in principle-- limiting cases corresponding to a vanishing dissipation rate or coupling with a thermal bath, respectively. It appears nonetheless that thermal behavior may emerge in this setting more often, as long as sufficiently strong dissipation is present. Some recent results on this emergence of classical critical behavior were obtained regarding the phase of free exciton-polariton condensates \cite{Kulczykowski2017,Comarondynamical2018, Sieberer2013,FF2017bistability} and the Dicke model \cite{Klinder2015,Lang2016,Maghrebi2019}, and include also non-markovian extensions \cite{nonmarkovianthermalization}.
Nevertheless, a number of genuinely non-equilibrium phase transitions are also known to be possible in this context \cite{Marcuzzi_2015,Diehl2016,Strack2015,Young2019,Gelhausen2018}.

One of the promising platforms in the study of dissipative phase transitions are arrays of quadratically (two-photon) driven Kerr cavities, objects originally suggested in the quest for noise-resilient quantum codes \cite{exacsolutionSciRep,Goto16,Goto2016,Nigg17,Puri2017} and for which building blocks have currently been experimentally realized with superconducting circuits \cite{Leghtas15,Wang16}. In this system, it has been explicitly demonstrated numerically that its symmetry-breaking phase transition \cite{VincenzoMF} exhibits a crossover from quantum- to classical critical behavior regarding the steady state properties for increasing dissipation in the form of single-photon losses. In particular, the system was found to belong to the universality class of the quantum or thermal Ising model respectively \cite{PRLspinmodel,twophotonGTA}. The crossover between these two spin models has also been subject to more general recent studies on the dynamics \cite{Foss_Feig_2013a,*Foss_Feig_2013b} and thermalization mechanism \cite{Jaschke_2019}.

Here, we are concerned with extending the conclusions in the large-dissipation limit to the dynamical aspects of the phase transition. In particular, we are interested in the dynamical critical exponent $z$ that relates the correlation time, and in closed quantum systems the Hamiltonian gap, to the control parameter and the correlation length. Near a critical point, the correlation time diverges, leading to the occurrence of critical slowing down. Some current works have addressed the latter effect in a dissipative context for optical bistability (a \emph{first-order} dissipative phase transition) \cite{Casteels16KZ,Vincentini18,Fink2018}, fermionic lattices \cite{electronslowingdown}, optical lattice clocks \cite{Changlatticeclock} and miscellaneous spin lattices \cite{RiccardoXYZ,Cai2013}.

For our study of the quadratically driven Kerr lattice, we will use two separate approaches. First, we address the occurrence of the Kibble-Zurek (KZ) effect. Because of the critical slowing down, domains are formed when the control parameter (photon driving in our case) is tuned through the transition at finite speed. According to a dynamical scaling hypothesis based on the KZ effect \cite{kibble_2007,scalingPolk}, we are able to extract $z$ from square lattices subject to a linear quench and obtain $1.9<z<2.3$.
Secondly, we study the slow relaxation dynamics to the steady state. This latter timescale is known to correspond directly to the inverse Liouvillian gap \cite{liouvillianSpectral}.
We here find scaling with $1.8<z<2.3$. The fact that both values are consistent implies that the Liouvillian gap exhibits the same scaling relations as the Hamiltonian gap  in closed quantum systems. 

The two-dimensional classical Ising model is known as a paradigmatic example of the second-order phase transition, both in terms of its historical importance as for its pedagogical value \cite{Ising1D,Ising2D,KersonHuang_2014}. In numerical simulations, it is typically implemented with a somewhat heuristic update rule, which is not uniquely defined by the steady state properties. Moreover, different update rules applied to the same model result in different values of $z$. The most well-known of these rules is the single-site update rule or Metropolis algorithm, which is consistent with experimental observations of the critical dynamics in an iron film \cite{Dunlavy2005}. The same value of $z$ is found  in the Hohenberg-Halperin model A dynamics \cite{HohenbergHalperin} as well as in $\phi^4$-theory \cite{Zhong2018}. 

We find here that the value of $z$  extracted from quadratically-driven Kerr resonators is indeed consistent with the value $z_m\approx 2.18$ from the Metropolis algorithm and model A dynamics. This correspondence between our nonequilibrium critical dynamics and model A dynamics is in line with Ref. \cite{FF2017bistability} on critical dynamics of a driven-dissipative Bose-Hubbard model.

\section{Quadratically driven photonic lattices}

The Hamiltonian of a $d$-dimensional lattice of nonlinear bosonic resonators with nearest-neighbour hopping and coherent two-photon driving, rotating at half the driving frequency, is
\begin{align}\label{eq:Hamiltonian}
    \hat{H}=&\sum_{i=1}-\Delta\cop_i\aop_i+\frac{U}{2}\aop_i^{\dagger 2}\aop_i^2+\frac{G}{2}{\aop_i}^{\dagger 2}+\frac{G^*}{2}\aop_i^2 \nonumber\\
    &-\sum_{\langle i j\rangle}\frac{J}{2d}(\cop_i\aop_j+\cop_j\aop_i),
\end{align}
where $\aop_i~(\cop_i)$ annihilates (creates) a photon at site $i$ and the last summation runs over nearest-neighbour pairs \cite{VincenzoMF}. $G$ is the two-photon driving amplitude, $U$ the Kerr nonlinearity, $J$ the hopping strength and $\Delta$ the detuning between half the driving frequency and the cavity frequency.
Furthermore, dissipation in the form of single-photon losses can, under Born-Markov approximation, be described with a jump operator $\sqrt{\gamma}\aop$, leading to a Lindblad master equation

\begin{equation}
 \label{eq:Liouvillian}
    \pdv{\hat{\rho}}{t}=\mathcal{L}\hat{\rho},=-i\comm{\hat{H}}{\hat{\rho}}+\gamma\sum_j\aop_j\hat{\rho}\cop_j-\frac{\gamma}{2}\acomm{\cop_j\aop_j}{\hat{\rho}}.
\end{equation}
Generally, two-photon losses can also be present in this system, but as suggested by analytic \cite{PRLspinmodel} and numeric \cite{twophotonGTA} arguments, these have little qualitative influence on the critical behavior \footnote{at least regarding the steady-state properties} while being computationally constraining, so that we don't include this process explicitly in this work.

A first study on system \eqref{eq:Liouvillian} was performed on the mean-field level in \cite{VincenzoMF} and predicted the occurrence of spontaneous breaking of the $\mathbb{Z}_2$-symmetry when increasing $G$. Recently, this prediction has been confirmed in a similar parameter regime with the additional finding that the transition belongs to the universality class of the thermal Ising model \cite{twophotonGTA}. In the weak-loss limit, by contrast, the transition belongs to the quantum-Ising universality class and the system can be explicitly mapped on a $XY$-spin model \cite{PRLspinmodel}. For negative $J$, antiferromagnetic behavior emerges \cite{antiferromagnet}. A brief look at the single-site problem \cite{exacsolutionSciRep} provides a simple picture to understand these behaviors: in each individual site, there are two metastable coherent state solutions with displacement $\pm\alpha_0$ (with a purely imaginary value in our considered case in absence of two-photon losses), that can be easily interpreted as $\ket{\uparrow}$ and $\ket{\downarrow}$ spin states. Alignment of these coherent states in neighbouring sites then corresponds to ferromagnetism, whereas a more random distribution $+\alpha_0$ and $-\alpha_0$ corresponds to classical paramagnetic behavior and maximal disalignment means antiferromagnetism. In the quantum paramagnet, the sites are highly entangled into superpositions of the coherent states.

The former studies were all restricted to steady-state properties of the transition. Here, we extend the results of Ref. \cite{twophotonGTA} obtained with the Gaussian Trajectory Approach (GTA) to the dynamical properties of the phase transition.

According to the quantum trajectory framework \cite{breuer,CarmichaelBOOK}, an open quantum system can be described as a stochastic average over quantum trajectories $\{\ket{\psi}_s\}$, corresponding to single-shot experimental realizations. Crucially, even though the full evolution \eqref{eq:Liouvillian} does not break the symmetry, individual trajectories $\{\ket{\psi}_s\}$ can. The GTA \cite{gaussianmethod,photoncondensate} adds a Gaussian ansatz to this formalism. This means that every trajectory $\ket{\psi}_s$ of an $N$-mode system is characterized only by the the coherent displacements $\{\alpha_{s,i}\}$ and the anomalous $\{u_{s,ij}\}$ and normal $\{v_{s,ij}\}$ quantum correlations where $1\leq i,j\leq N$. Explicitly, these coefficients are defined as 

\begin{align}
\alpha_{s,i}&=\bra{\psi_s}\aop_i\ket{\psi_s}\nonumber\\
u_{s,ij}&=\bra{\psi_s}\aop_i\aop_j\ket{\psi_s}-\alpha_{s,i}\alpha_{s,j}\nonumber\\
v_{s,ij}&=\bra{\psi_s}\cop_i\aop_j\ket{\psi_s}-\alpha_{s,i}^*\alpha_{s,j}.
\end{align}

The Gaussian ansatz thus reduces the complexity of each trajectory to quadratic as function of system size. In ref. \cite{twophotonGTA}, the corresponding GTA equations were derived explicitly for the quadratically driven photonic lattice.
By solving these to the steady state, emergence of an ordered phase was first witnessed by a macroscopic occupation of the $k_0$-mode. Furthermore, the quantity $\ol{\alpha} = \Im \frac{1}{N}\sum_i\alpha_i$ becomes a suitable real-valued order parameter akin to a magnetization of which the distribution (sampled by values $\ol{\alpha}_s$) changes from monomodal in the paramagnetic (disordered) phase to bimodal in the ferromagnetic (ordered) phase. Using finite-size scaling of the Binder cumulant, which quantifies this behavior \cite{Binder1981}, the critical exponent $\nu=1$ was extracted, indicating that the transition belongs to the universality class of the classical Ising model; as well as the critical value $G_c\approx0.86$ for the parameters $U=\gamma=J=1,\Delta=-1$. 

For the dynamical numerical studies in this work, we are able to evolve these GTA equations with the same timestep as the static case \cite{twophotonGTA}, ($h=10^{-4}$ for the Euler-Maruyama method which coincides with Milstein's method \cite{milstein_tretyakov_2010}) because our interest is focused on timescales that are slower than the ones corresponding to individual Hamiltonian or dissipation terms.

\section{Kibble-Zurek scaling at a linear quench}

When a parameter in a thermodynamic system is slowly varied, the adiabatic theorem assures that the system remains at all times in an equilibrium state at constant entropy. The minimal ramp time for equilibrium to be preserved is given by the relaxation time. When a parameter is quenched through a second order phase transition, the relaxation time diverges, adiabaticity always breaks down and domains with different values of the symmetry-breaking order parameter are formed. 
This mechanism is known as the Kibble-Zurek (KZ) mechanism \cite{kibble_2007}. Originally introduced in a cosmological context \cite{Kibble1980}, it has been developed further mainly in condensed matter systems, with first applications to liquid helium \cite{Zurek1985} and rotor models, where the defects are vortices \cite{kibble_2007}. More recently, the KZ mechanism has been applied to more generic systems where the defects take the form of domain walls \cite{Chandran2012,Sabbatini_2012}. Extensions to quantum phase transitions have also been developed \cite{Dziarmaga2010,rmpPolkovnikov,scalingPolk,Jaschke_2017,Silvi2016}.

\begin{figure}
    \centering
    \includegraphics[width=0.9\linewidth]{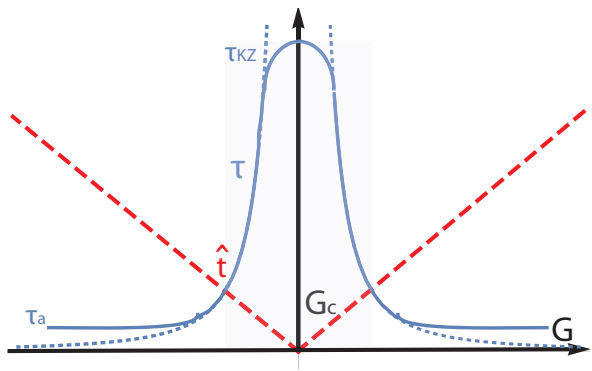}
    \caption{The Kibble-Zurek effect: when $G$ is linearly increased, the time until $G_c$ is reached (red, dashed line) becomes less than the diverging correlation time $\tau$ (blue line) at $\hat{t}$. From $\hat{t}$ onwards, the dynamics freezes (blue region). In a finite system, the true value of $\tau$ (full line) only follows the universal behavior for intermediate velocities $v$ for which the crossing occurs at $\tau_a<\tau<\tau_{KZ}$.}
    \label{fig:KZ}
\end{figure}

\begin{figure}
    \centering
    \includegraphics[width=0.9\linewidth]{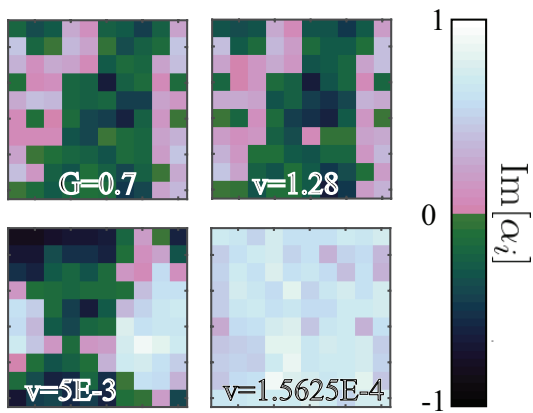}
    \caption{During the quench protocol, $G$ is increased linearly with time from $G_0=0.7J$ to $G_c=0.86J$, the critical point \cite{twophotonGTA}. The upper left panel shows a Monte Carlo sampling of the steady state at the $G=G_0$. The other panels show a sample of the final state that was evolved with different quench speeds $v$ For $v\gtrsim 1$, almost no evolution has been able to take place. For decreasing $v$, the correlations are able to spread further, until for $v\lesssim v_{kz}(L)$ the whole 10x10 lattice is correlated. (color online)}
    \label{fig:islands}
\end{figure}

More specifically, the KZ mechanism, works as follows (Fig. \ref{fig:KZ}). We envision a continuous quench where, starting from the steady-state solution at some value $G_0$ in the paramagnetic regime, $G$ is increased linearly up to $G_c$ in a total time $T$:
\begin{equation}
G(t)=G_0+vt\qquad 0\leq t \leq T,
\end{equation}
where $v=\frac{G_c-G_0}{T}$ .
As $G_c$ is approached, the correlation time  diverges as  $\tau\sim (G-G_c)^{-z\nu}$ while the correlation length scales as $\xi\sim (G-G_c)^{-\nu}$.  From some point in time during the quench $T-\hat{t}$ onward, $\tau(G)>\hat{t}$:  the dynamics freezes as the time available for the dynamics drops below the correlation time.  The correlation length is then unable to increase beyond $\xi_c\sim (G(T-\hat{t})-G_c)^{-\nu}$, setting the final domain size. Crucially, the lower $v$, the larger $G(T-\hat{t})$ and thus the larger the domains. The former arguments assume universality in an infinite lattice. An example of such domain formation after quenching with different $v$ from a given initial state is given in Fig. \ref{fig:islands}.

Focussing on the  steady-state properties of a more realistic finite lattice system, there are three distinct length scales present: the lattice spacing $a=1$, correlation length $\xi$ and system size $L$. Close to criticality, one has $\xi\gtrsim a$ leading to the scaling hypothesis that all quantities can be expressed as a function of the dimensionless ratio $\xi/L$ \cite{cardyfinitesize}. Such approach was also used to study system \eqref{eq:Liouvillian} in previous work \cite{twophotonGTA}.

For dynamic scaling, we now follow the scaling approach introduced in ref. \cite{scalingPolk}.
 Three different timescales are present: $\tau_a$ relates to the local individual Hamiltonian processes and loss rate, $\tau$ the correlation time, and $\tau_{KZ}$ the time scale for the adiabatic evolution of the finite system. In general, universal behavior for $\xi$ and $\tau$ , and hence standard Kibble-Zurek scaling, can be expected for $\tau_a<\hat{t}<\tau_{KZ}$ (Fig. \ref{fig:KZ}).
Each of these timescales is associated with velocity scales \cite{scalingPolk}.

On the order of the lattice spacing \footnote{More precisely, $v_a$ relates to the energy scales of individual Hamiltonian processes and dissipation rate, which we all take of order one.},
 $v_a\sim J^2a^{-(z+1/\nu)}=J^2$  marks the transition speed between possibly microscopic dynamics and long-range universal behavior. The second scale is the quench speed $v$. Third, there is  the Kibble-Zurek speed $v_{KZ}(L)\sim J^2 L^{-(z+1/\nu)}$: marking the speed where domain size becomes comparable with the lattice size (below $v_ {KZ}(L)$ the dynamics remain adiabatic).
 Now, two separate scaling functions appear for the second moment of the order parameter after the quench $(t=T)$ in different regimes \cite{scalingPolk}. If the quench is sufficiently slow for the microscopic processes to be unimportant $(v\lesssim v_a)$, one expects universal scaling as a function of $v/v_{KZ}$:

\begin{equation}
    \ev{\ol{\alpha}^2}=L^{-2\beta/\nu}f_1(vL^{z+1/\nu}),
\end{equation}
where $\ev{\cdot}$ denotes a statistical expectation value over trajectories. On the other hand, if the quench is sufficiently fast for the finite size to be unimporant ($v\gtrsim v_{KZ}(L)$), universal scaling of $v/v_a$ is expected, leading to
\begin{equation}
    \ev{\ol{\alpha}^2}=L^{-d}f_2(v^{-1}).
\end{equation}
In the overlapping region $v_{kz}\lesssim v \lesssim v_a$, both scaling functions overlap with a power-law dependence $\ev{\ol{\alpha}^2}\propto v^{-x}$ with exponent
\begin{equation}\label{eq:powerlaw}
    x=\frac{d-2\beta/\nu}{z+1/\nu}.
\end{equation}

The above arguments have extensively been verified for thermal systems, but the generic nature of arguments suggests that they should also work out of equilibrium. In order to verify the KZ mechanism in the two-photon driven dissipative Hubbard model, we have simulated quenches in the amplitude of the two-photon drive of the Bose-Hubbard model from $G_0=0.7J$ to $G_c=0.86J$.

\begin{figure*}
\includegraphics[width=0.9\linewidth]{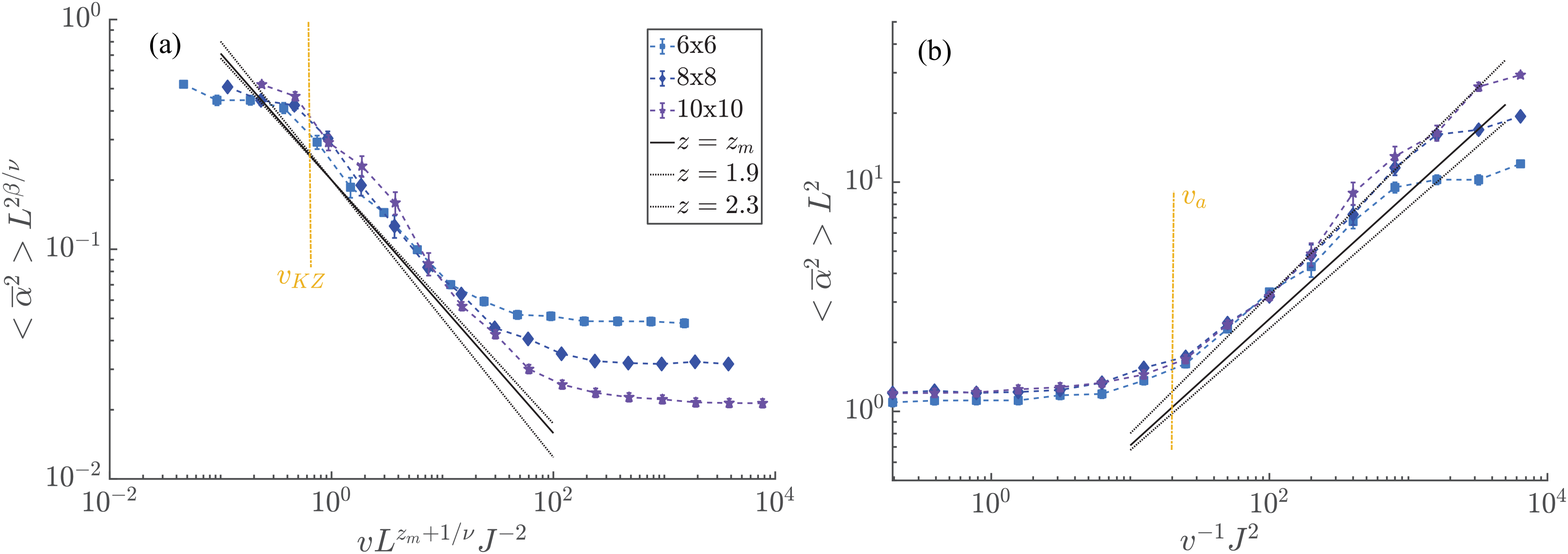}
\caption{Dynamic scaling functions for different system sizes showing collapse in the adiabatic  ($f_1$,(a)) and diabatic limit  ($f_2$,(b)), for linear quenches from $G_0=0.7J$ to $G_c=0.86J$. In the intermediate regime, there is a universal power-law scaling (black line) with exponent $x=\frac{d-2\beta/\nu}{z+1/\nu}$, where dimension $d=2$; $\beta=0.125,~\nu=1$ are critical exponents of the universality class of the 2D classical Ising model; and $z$ is the dynamical critical exponent. Results are consistent with $z=z_m\approx2.18$, the dynamical critical exponent characterizing Metropolis dynamics in the 2D Ising model,  this value was also used for the finite-size scaling itself. Parameters: $U=\gamma=J=1,\Delta=-1$. For the fastest quenches ($v/J^2\geq10^{-2}$), $10^3$ trajectories were used, and $10^2$ trajectories for the slower quenches. The yellow vertical lines denote approximate values for $v_{KZ}$ and $v_a$, marking the edges of regime where power-law scaling is valid. We find that the prefactor in the definition of $v_{KZ}$ is of order one, whereas $v_a\approx 10 J^2$}.
\label{fig:scaling}
\end{figure*}

In Fig. \ref{fig:scaling} the extracted slow and fast scaling functions $f_1$ and $f_2$ are shown for different square lattice sizes, and collapse is observed for both when using exponents $\beta=0.125,~\nu=1,~z\approx2.18$. The former two numbers are static critical exponents corresponding to the 2D thermal Ising model \cite{twophotonGTA}. The value of $z$, the dynamical critical exponent, corresponds to Metropolis dynamics in this model \footnote{The same classical model can have different simulation algorithms (update rules) with other values of $z$ \cite{scalingPolk}}. In the intermediate regime for $v$, the predicted power-law \eqref{eq:powerlaw} is further observed consistent with the aforementioned values of the critical exponents.

We thus have obtained strong evidence for the fact that not only the static, but also the dynamical properties of the  two-photon driven Bose-Hubbard model are in the Ising universality class, more precisely of Metropolis dynamics.

\section{The Liouvillian gap scales with the same exponent}

In Hamiltonian systems, the dynamical exponent $z$ is further known to govern the scaling of the gap $\Delta_H$ \cite{sachdev}:
\begin{equation}\label{eq:hamgapscaling}
\Delta_H\sim\xi^{-z}\text{   \&   }\Delta_H\sim\abs{g-g_c}^{z\nu},
\end{equation}
where $g$ is the control parameter. Because system \eqref{eq:Liouvillian} studied here belongs to a classical universality class due to its driven-dissipative nature, the fate of relations \eqref{eq:hamgapscaling} is not \emph{a priori} clear.
In an open system, the dynamics are governed by the Liouvillian superoperator $\mathcal{L}$ \cite{liouvillianSpectral}. Likewise, the timescale of slowest relaxation to the steady state is determined by the Liouvillian gap $\lambda$, defined as (minus) the real part of the first nonzero eigenvalue of $\mathcal{L}$. One can thus ask if $\Delta_H$ can be replaced by $\lambda$ in relations \eqref{eq:hamgapscaling}.
If  this replacement in the first relation of \eqref{eq:hamgapscaling} is valid, then one must have in a finite system \cite{cardyfinitesize}
\begin{equation}\label{eq:gapscaling}
    \lambda=\xi^{-z}\tilde{f}(\xi/L)=L^{-z}f(L^{1/\nu}(G-G_c)),
\end{equation}
where $\tilde{f},f$ are unknown scaling functions. 

\begin{figure*}
\includegraphics[width=0.9\linewidth]{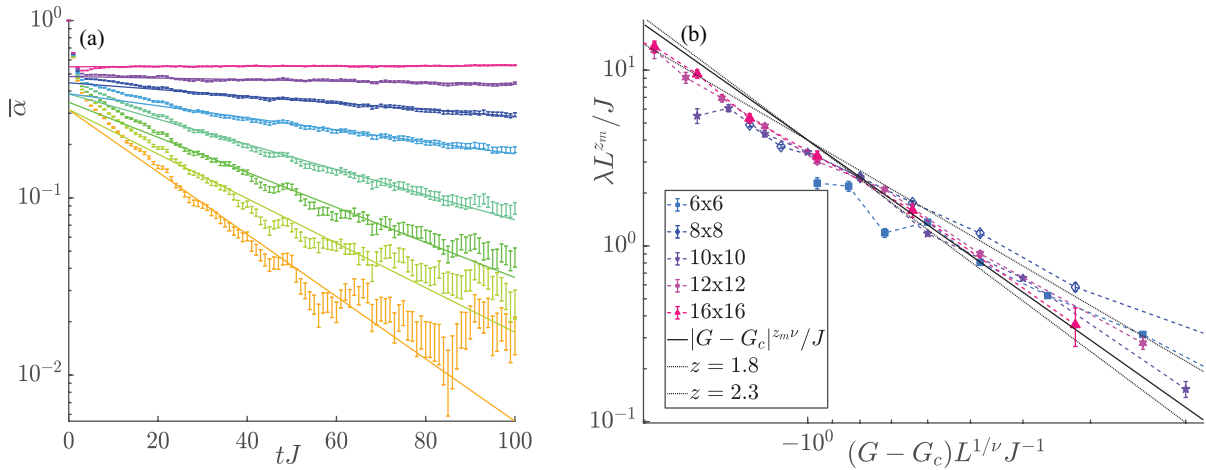}
\caption{(a): Errorbars show the decay of $\ol{\alpha}$ to the steady state value (0) in a 10x10 lattice after initialization in a polarized $\alpha_j=1i,\forall j$ state, average over trajectories. From bottom to top, $G$ values run increment from $0.74J$ with steps $0.02J$ to $0.86J$. Full lines in corresponding colors: fits $\propto e^{-\lambda t}$, with fitting from $tJ=20/30/40$ onwards (6x6,10x10/12x12/16x16) ($tJ=10-50$ for the 8x8 case). (b):  extracted values of $\lambda$, rescaled assuming $z=z_m$ as function of rescaled $G$. Especially  in the $\xi < L$ regime (where the function argument less than -1), collapse is very clear and in agreement with power-law behavior \eqref{eq:gapscaling2}. Closer to the critical point, the data collapse becomes worse. It should be noted that in this regime, the time scales become very long and the extracted decay rate may be less accurate. Moreover, the slow dynamics is more sensitive to rare events, that may not be sufficiently sampled. $10^3$ trajectories were used in the simulations up to 12x12 lattices and $10^2$ for the 16x16 case.}
\label{fig:gaps}
\end{figure*}

In order to obtain values of $\lambda$ for different values of $G$ and $L$ numerically, we perform the following procedure in each case. After starting from a fully polarized ($\alpha_j=1i,\forall j$) state, the system is left to evolve freely. An exponential $\sim e^{-\lambda t}$ is then fitted to the slow relaxation process towards the steady state, as illustrated for a 10x10 lattice in the left panel of Fig. \ref{fig:gaps}.


Validity of Eq. \eqref{eq:gapscaling} implies a collapse of curves for different system sizes when plotting  $L^z\lambda$ as function of $L^{1/\nu}(G-G_c)$. In the right panel of Fig. \ref{fig:gaps}, we see that this is indeed the case for $\nu=1,z=z_m$, the same values as in the previous section.

Furthermore, relation \eqref{eq:gapscaling} also directly relates $\lambda$ to control parameter $G$. In the limit of large system sizes, we observe indeed (black line on Fig. 2) 
\begin{equation}\label{eq:gapscaling2}
\lambda\sim\abs{G-G_c}^{z\nu},
\end{equation}
where the numerical result is especially good in the regime $\xi<L$.
This means that both relations \eqref{eq:hamgapscaling} are valid for the considered Liouvillian dynamics.

\section{Conclusions}

In a dissipative Bose-Hubbard model with two-photon driving, the classical Ising model can be simulated, where the role of magnetization is taken by polarization of the optical phase. Not only does the transition in this system belong to the universality class of the Ising model at equilibrium, but also the dynamical properties match as witnessed by the value of critical exponent $z$.

We have also shown that the Liouvillian gap $\lambda$ scales with the same exponent $z$, with relations very reminiscent of the scaling of Hamiltonian gaps at quantum phase transitions in closed systems. To what extent this scaling behavior is generic for open quantum systems is an interesting open problem.

It is further interesting to note that there are alternatives for metropolis in classical Ising simulations, with different values of $z$, which can converge to the steady state faster (Swedsen-Wang, Wolff) or slower (East). One may wonder if suitable reservoir engineering would allow simulation of these as well. This could be useful to speed up or slow down relaxation to the same steady state numerically or experimentally. A possibility would be exploiting the difference between one common bath or independent baths, which has shown to reflect at least on the decoherence time of an open quantum system \cite{Jaschke_2019}

It would also be interesting to see how the dynamical criticality behaves in the quantum regime, where Kibble-Zurek effect can exhibit richer behavior \cite{Silvi2016}, as has also been suggested to study experimentally with trapped ions \cite{Puebla2019}.
Recently, also anti-Kibble Zurek behavior was found under certain circumstances in open quantum systems \cite{Dutta2016,Puebla2019anti}.

Even within a classical regime, optical simulation of the Ising model has been proposed to solve NP-hard tasks \cite{Barahona_1982}, including through degenerate parametric oscillators \cite{Marandi2014,Inagaki603,McMahon614}, a system with an analogous symmetry breaking to ours. To this purpose, our results also point out that, especially in absence of an all-to-all connected setup, care must be taken when driving through the transition that $v<v_{KZ}$ in order to find the true steady state (Ising ground state) and not a metastable state with Kibble-Zurek domains.
Recently also implementations of the Ising model in a Kerr resonator have been proposed \cite{Kyriienko2019} where optical bistability in the single-photon driven case is used to map the two spin states. Unlike this situation, the $\mathbb{Z}_2$ symmetry is exact in Kerr resonators with two-photon driving studied here, which might benefit the accuracy of the results of such optimization algorithms.

\acknowledgments

We acknowledge stimulating discussions and comments on the manuscript from R. Rota and F. Minganti. The computational resources and services used in this work were provided by the VSC (Flemish Supercomputer Center), funded by the Research Foundation - Flanders (FWO) and the Flemish Government – department EWI. Financial support from the project FWO-39532 is acknowledged.


\providecommand{\noopsort}[1]{}\providecommand{\singleletter}[1]{#1}%

\end{document}